%% file: main.tex
\documentclass[sigconf]{acmart} 
\AtBeginDocument{%
  }

\usepackage{subcaption}

\usepackage{xfrac}

\newcommand{\mqss}{\gls{mqss}}
\newcommand{\lrz}{LRZ}

\setcopyright{acmlicensed}
\copyrightyear{2025}
\acmYear{2025}
\acmDOI{10.1145/3731599.3767551}
\acmConference[SFWM]{International Workshop on Software Frameworks and Workload Management for Quantum and HPC Ecosystems}{November 21, 2025}{St. Louis, MO}
\acmISBN{979-8-4007-1871-7/2025/11}

\begin{document}

\title[First Practical Experiences Integrating Quantum Computers with HPC Resources]{First Practical Experiences Integrating Quantum Computers with HPC Resources: A Case Study With a 20-qubit Superconducting Quantum Computer}


\author{Eric Mansfield}
\email{eric.mansfield@meetiqm.com}
\affiliation{%
  \institution{IQM Quantum Computers}
  \city{Munich}
  \state{Bavaria}
  \country{Germany}
}

\author{Stefan Seegerer}
\email{stefan.seegerer@meetiqm.com}
\affiliation{%
  \institution{IQM Quantum Computers}
  \city{Munich}
  \state{Bavaria}
  \country{Germany}
}

\author{Panu Vesanen}
\email{panu.vesanen@meetiqm.com}
\affiliation{%
  \institution{IQM Quantum Computers}
  \city{Espoo}
  \state{Uusimaa}
  \country{Finland}
}

\author{Jorge Echavarria}
\email{jorge.echavarria@lrz.com}
\affiliation{%
  \institution{Leibniz Supercomputing Centre}
  \city{Garching}
  \state{Bavaria}
  \country{Germany}
}

\author{Burak Mete}
\email{burak.mete@lrz.com}
\affiliation{%
  \institution{Leibniz Supercomputing Centre}
  \city{Garching}
  \state{Bavaria}
  \country{Germany}
}

\author{Muhammad Nufail Farooqi}
\email{muhammad.farooqi@lrz.de}
\affiliation{%
  \institution{Leibniz Supercomputing Centre}
  \city{Garching}
  \state{Bavaria}
  \country{Germany}
}

\author{Laura Schulz}
\email{schulz@anl.gov}
\affiliation{%
  \institution{Argonne National Lab}
  \city{Lemont}
  \state{Illinois}
  \country{USA}
}

\renewcommand{\shortauthors}{Mansfield et al.}

\begin{abstract}
  \input{src/00_abstract}
\end{abstract}

\ccsdesc[500]{Hardware~Quantum computation}
\ccsdesc[300]{Applied computing~Data centers}
\ccsdesc[300]{Social and professional topics~Computer manufacturing}
\ccsdesc[300]{Social and professional topics~Hardware selection}
\ccsdesc[500]{Social and professional topics~Management of computing and information systems}
\keywords{Quantum computing, high-performance computing, hybrid workflows, superconducting qubits, HPC+QC Integration, HPCQC, site selection}



\maketitle

\input{src/01_intro}
\input{src/02_space}
\input{src/03_daily_ops}

\input{src/04_onboarding}

\input{src/05_conclusion}

\input{src/06_acks}

\bibliographystyle{ACM-Reference-Format}
\bibliography{bib}

\end{document}

%% file: src/00_abstract.tex
Incorporating Quantum Computers into High Performance Computing (HPC) environments (commonly referred to as HPC+QC integration) marks a pivotal step in advancing computational capabilities for scientific research. Here we report the integration of a superconducting 20-qubit quantum computer into the HPC infrastructure at Leibniz Supercomputing Centre (\lrz), one of the first practical implementations of its kind. This yielded four key lessons: (1) quantum computers have stricter facility requirements than classical systems, yet their deployment in HPC environments is feasible when preceded by a rigorous site survey to ensure compliance; (2) quantum computers are inherently dynamic systems that require regular recalibration that is automatic and controllable by the HPC scheduler; (3) redundant power and cooling infrastructure is essential; and (4) effective hands-on onboarding should be provided for both quantum experts and new users. The identified conclusions provide a roadmap to guide future HPC center integrations.

%% file: src/01_intro.tex
\section{Introduction}

Quantum computers, leveraging the principles of quantum mechanics, offer unique capabilities for solving specific problems that are intractable for classical systems. These include chemistry simulation, combinatorial optimization, and machine learning, where quantum systems could provide exponential speedups or energy efficient solutions. Classical computing, particularly High Performance Computing (HPC), is well-suited for parallelized problems with large datasets (“big data”). Quantum Computing can complement the capabilities of HPC by enabling specific High Performance Computing+Quantum Computing (HPC+QC) workflows, a series of steps, that combine the unique strengths of both classical and quantum computing to create new methods for scientific computing \cite{humble2021}. 
However, the practical integration of quantum computers into HPC workflows remains a significant challenge due to their unique operational requirements and the need for seamless interaction with classical resources \cite{beck2024integrating}.

Quantum computers are a very young technology and only entered the first HPC centers around 2024. Most quantum computers resemble experimental lab setups that were exclusively built in academic research labs until only a few years ago and do not fit into a compact form factor like HPC nodes. The unusual shape and footprint precludes quantum computers from being placed in line with conventional HPC nodes. In addition, while HPC nodes are delivered mostly pre-assembled needing mainly connection to electricity and cooling, quantum computers are often assembled on site by a team with expertise in cryogenics, superconductors, and microwave components or optics depending on the exact system. Their daily operations differ as well; while HPC nodes have very constant properties that are determined in fabrication, quantum computers are dynamic systems with changeable properties that must be managed via regular calibration.

The second major challenge is the software integration of on-premise quantum computers within the existing HPC workflow orchestration system. For quantum computers to be an effective accelerator, it is necessary for them to have high levels of availability and uptime, be able to schedule and manage jobs within the existing resource management frameworks, and to minimize the amount of time the quantum computers spend doing calibration. 

There are many modalities of quantum computers today (superconducting, neutral atoms, trapped ions, photonics, etc). Of these, the supercomputing center selected a quantum computer based on superconducting transmon qubits. The transmon qubit \cite{koch2007charge}, is one of the most mature quantum computing modalities with close to 20 years of continuous innovation. In this paper, we recount the practical experience in installing a 20 qubit quantum computer - detailing from selecting of the site, the utilities required, installation, and daily operations of a quantum computer in an HPC center.

The computer being installed is an on-premise superconducting quantum computer targeted to R\&D laboratories and HPC centers for use case exploration and research purposes  \cite{abdurakhimov2024technology}. The full stack solution includes the Quantum Processing Unit (QPU), the control electronics, the cryogenic system, and a full software stack. The QPU has 20 superconducting transmon qubits in a square grid topology, where the tunable couplers mediate the connection between each qubit pair. The software interfaces allow users to write their experiments and algorithms using widely known frontend frameworks such as Qrisp\cite{qrisp}, Qiskit\cite{qiskit} and Cirq\cite{cirq}, or as pulses.

\begin{figure*}[ht]
    \centering
    \includegraphics[width=0.86\textwidth]{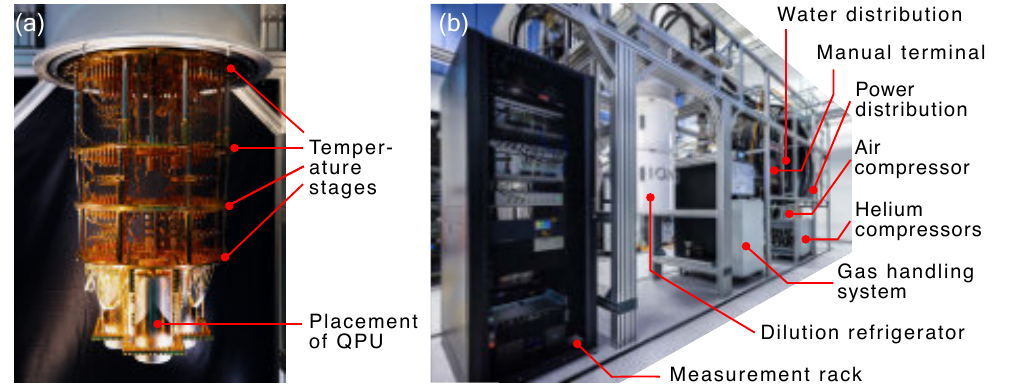}
    \caption{a) The inner parts of the white cylinder containing the QPU and a tiered temperature stages assembly (more commonly known as the “chandelier”). This image shows a similar setup. (b) Photograph of the system without covers.}
    \label{chandellier}
    \Description{Photograph showing the inner parts of a white cylinder containing the QPU and a tiered temperature stages assembly, also known as the chandelier, and the system without covers.}
\end{figure*}

Designed with HPC integration in mind, its overall dimensions are close to a row of HPC nodes (126 cm wide x 453 cm long x 290 cm tall). It consists of several main components (see figure \ref{chandellier}). The most visible part is the cryostat, the white cylinder containing the QPU and a tiered temperature stages assembly (more commonly known as the “chandelier”). The cryostat is under vacuum during operation and cooled in stages down to $10$ mK. A standard 19" rack contains the room temperature control electronics that generate and receive microwave signals controlling the superconducting qubits. The rack also contains a small classical computer running a Linux distribution for controlling the overall system. The other main components are the cryogenic cooling system consisting of a gas handling system driven by turbomolecular pumps to circulate low pressure gaseous helium into the cryostat and a compressor supplying air to pneumatic valves to control the flow of the gases.

The following sections detail the practical steps taken and the lessons learned in integrating a quantum computer into a high-performance computing center. The system discussed represents a "co-located loose integration" model \cite{mete}, a foundational step toward future tighter integration. We address the hardware and environmental prerequisites necessary to host a quantum computer and the process of its installation and commissioning (Chapter \ref{02_space}), the daily operational model designed for stability and remote management (Chapter \ref{03_daily_ops}), and the crucial steps for enabling users to build effective hybrid quantum-classical workflows (Chapter \ref{04_onboarding}).

%% file: src/02_space.tex
\section{Selecting the Space and Installation}
\label{02_space}

The first step in deploying the quantum computer is to determine a suitable location within the facility to house the system. One of the key differences between superconducting quantum computers and classical HPC nodes is an increased sensitivity to noise. This primarily comes from two sources: electronic/magnetic noise and acoustic/vibrational noise. These sources of noise can produce a range of deleterious effects including degrading the coherence time of the qubits (time that the information is retained in its quantized and entangled state), distorting gate execution and corrupting the information stored in qubits \cite{PRXQuantum.3.040332}. 
Another important difference to traditional HPC nodes, where signals are primarily carried \textit{in silico} or routed on printed circuit boards, superconducting qubits utilize dozens to hundreds of fine gauge cables to transmit and receive microwave signals. External sources of vibration or acoustic noise can cause these cables to vibrate, introducing triboelectric noise and electromotive noise \cite{Kalra_2016, kono2024mechanically}. In practice, sources of noise including trains or trams running near the building, air conditioning chillers and Finnish death metal played at high volume have been seen. 
Accordingly, the space selection process must address both these unique environmental sensitivities and the conventional requirements of a compute resource, namely the provision of stable power delivery, the integration with the facility's cooling infrastructure, the setup of network connections for system control, and the software integration.

\subsection{Site survey}
A thorough site survey and preparation phase was therefore the first critical step in ensuring the system can operate reliably with minimal disruption. From prior experience, it was known that the presence of tram lines (“streetcars”), underground lines (“subways” or “metros”), or other heavy traffic (e.g. the Autobahn) can cause problematic vibrations. In addition, the site should be at least $100$ m from cellular base stations, radio transmitters and other sources of high energy non-ionizing radiation. Fluorescent lighting is also a known source of noise and must be at least $2$ m from the system.

The HPC center selected three potential spaces to house the quantum computer following these rough guidelines. Then engineers from the quantum computer manufacturer went on site to measure the environmental conditions in a site survey. This included measurements of DC and AC magnetic fields, floor vibrations, temperature, humidity, and sound pressure level. The duration of temperature and humidity measurement needed to be at least $25$ hours to capture a full cycle of typical building conditions. Suitable measurement equipment and appropriate acceptance limits are given in Table \ref{tab:measurements}.  

The other critical consideration that led to the final selection of the space was the access path to physically deliver and install the quantum computer. The complete delivery path from loading dock, elevators, hallways, doorways, and to a small staging area in the space for the system, all needed a minimum width of $90$ centimeters. Lastly, the 20-qubit quantum computer is more lightweight than a state-of-the-art HPC cluster and requires the floor to support only $1000~ \sfrac{\text{kg}}{\text{m}^2}$ ($205 \sfrac{\text{lbs}}{\text{ft}^2}$). Thus, floor strength was not a limiting factor in selecting the location.

\begin{table*}[ht]
\centering
\begin{tabular}{|p{2.7cm}|p{6.8cm}|p{6.6cm}|}
\hline
\textbf{Measurement} & \textbf{Equipment} & \textbf{Requirement} \\
\hline
DC magnetic field & A 3-axis fluxgate sensor positioned at the planned location of the cryostat, approximately at the height of the QPU. & $< 100$ µT for each of the axes. \\
\hline
AC magnetic field & A 3-axis fluxgate sensor positioned at the planned location of the cryostat, approximately at the height of the QPU. & $<$ 1 µT peak-to-peak spectrum amplitude for each of the axes at frequency range $5$ Hz - $1000$ Hz. \\
\hline
Floor vibrations & A single axis vibration sensor positioned on the floor at the planned location of the cryostat. & $< 400 ~\sfrac{\text{µm}}{\text{s}}$ RMS spectrum amplitude for each of the axes at frequency range $1$ Hz - $200$ Hz, or equivalently, ISO vibration limit for office spaces. \\
\hline
Sound pressure & An omnidirectional microphone positioned at the location of the cryostat. & $< 80$ dBA when integrated over the frequency range $20$ Hz – $20$ kHz. \\
\hline
Temperature & A thermometer positioned at the planned location of the electronics cabinet. & $\Delta T < \pm 1^\circ\text{C}$ within 12 hours around any set point between $20 - 25 ^\circ\text{C}$. \\
\hline
Humidity & A hygrometer positioned at the planned location of the electronics cabinet. & $25 - 60\%$, non-condensing \\
\hline
\end{tabular}
\vspace{0.5em}
\caption{The measurement equipment and acceptance criteria for the site survey measurements.}
\label{tab:measurements}
\end{table*}

\subsection{Power Consumption}
A superconducting quantum computer with 20 qubits consumes power from three sources: electrical power to run its control electronics and cryogenic gas handling system, room air conditioning for extracting heat from the control electronics and cooling water for extracting heat from the cryogenic cooling system. In total, the superconducting quantum computer uses only modest amounts of power with a peak power consumption of $30$ kW during cooldown to operating temperatures.

In comparison, a classical HPC node Cray EX4000 cabinet can draw up to $141$ kVA ($\sim 140$ kW real power) under standard configurations according to HPE's official specifications \cite{hpe}. The associated Cray EX cooling infrastructure supports up to $1.2$ MW for four cabinets, implying a per‑cabinet power capability of approximately $300$ kW in high‑density scenarios. Thus existing HPC centers will have sufficient electrical power capacity for deploying superconducting quantum computers.

\subsection{Cooling Water}
Cooling water is needed for superconducting quantum computers for extracting heat from the cryogenic cooling components: the pulse tube compressor(s) and the turbomolecular pumps of the gas handling system. These components were originally designed for use in academic research labs and their requirements are not identical to those of water cooled HPC racks. Many HPC racks accept incoming cooling water with temperatures of up to 45°C, whereas the cryostat manufacturer recommends a water temperature between 15 and 25°C.

The room temperature microwave control electronics cabinets have no active cooling and release their heat to the surrounding air, where the heat needs to be eventually removed by the facility air conditioning system. In addition to removing heat from the room temperature control electronics, air conditioning is also critical for maintaining a stable calibration. Superconducting quantum computers have microwave readout lines connecting the cryostat and control electronics. Small changes in ambient temperature  can cause phase delay in cabling and electronics, affecting the readout signals. Experience has thus shown that it is ideal to keep the ambient temperature change to $\Delta T < 1$ °C per 24 hours. A value that was achievable in practice.

\subsection{Network Connections}
The data transfer needs of near-term quantum computers are modest compared to leading HPC systems. The 20 qubit system described in this work was connected via $1$ Gbit ethernet to HPC  resources. Data transfer occurs in a few different steps while running a quantum computation, but the dominant bandwidth bottleneck is the output of a quantum job. The output is transferred from QPU to classical resources, which depending on the application can have one of several formats. The most common output format for circuit-based jobs is a histogram of the measured bitstrings and the number of their observed occurrences, where each element of the bitstring describes the binary-classified readout result of an individual qubit. Arranging the data in histograms of sampled bitstrings may also reduce data transmission requirements, depending on the structure of the state expected to be measured. Alternatively, in a more generic case, the output may simply be the individual bitstrings for each of the prescribed shots. For pulse-level experiments and readout-related work, the output can also be the raw complex (a + b\textit{i}) readout data represented as a pair of floating-point numbers for each of the qubits and each of the measured shots. 

To estimate the necessary bandwidth, consider the following naïve calculation. This system has its outputs represented as bitstrings for each of the measured shots. First, we assume that the measured circuits use a passive waiting period of duration 300 µs at the beginning of each shot to initialize the qubits to their ground states. This passive reset will dominate the duration of each of the shots. Second, we will assume that data is collected from all 20 qubits. Third, we will assume that the data format used for data transmission has some inefficiency, so that the representation of each measured bit consumes 8 bits (compared to a single bit in an ideal scenario). With these assumptions, a continuous measurement of circuits results in data rate of 1/300 µs x 20 x 8 bit = 533 kbit/s, which is well below the transmission rate offered by the 1 Gbit Ethernet connection used in this work. In practice, the control software has additional inefficiency, so that fully continuous measurements are not possible, further reducing the network bandwidth needs. Extending the above calculation from 20 to 54 or 150 qubits shows that the data rate grows linearly as the number of qubits increases.


\subsection{Physical Installation}
While traditional HPC nodes are delivered partly pre-assembled racks that can be wheeled into place, connected to both electricity and cooling, and then commissioned, quantum computers are often assembled on site, as was done in this project. 

The multi-day (or multi-week) process of assembly requires bringing components in large wooden crates (the cryostat of a superconducting quantum computer weighs approximately 750 kg), testing hundreds of factory connected microwave signal lines and ultimately assembling everything within a production environment.

\subsection{Software integration to HPC resources}

HPC+QC integration also requires a software architecture capable of accommodating two fundamentally distinct user-interaction modes. The first mode relies on remote, Application Programming Interace (API)-based asynchronous access: users submit jobs to a queue which are later executed on a QPU, a mode better suited for workloads with minimal classical-quantum interaction. The second mode involves treating the QPU as an accelerator in a classical HPC workflow, allowing quantum operations to be executed within a tightly-coupled, low-latency loop. Such a model is essential for hybrid quantum-classical algorithms such as the Variational Quantum Eigensolver (VQE) \cite{cerezo2021variational}. 

Crucially, both paradigms demand a compiler toolchain that translates high-level quantum code into low-level QPU instructions by leveraging a common Intermediate Representation (IR), enabling homogeneous compilation strategies across heterogeneous targets.

\begin{figure*}[ht]
    \centering
    \includegraphics[width=\textwidth]{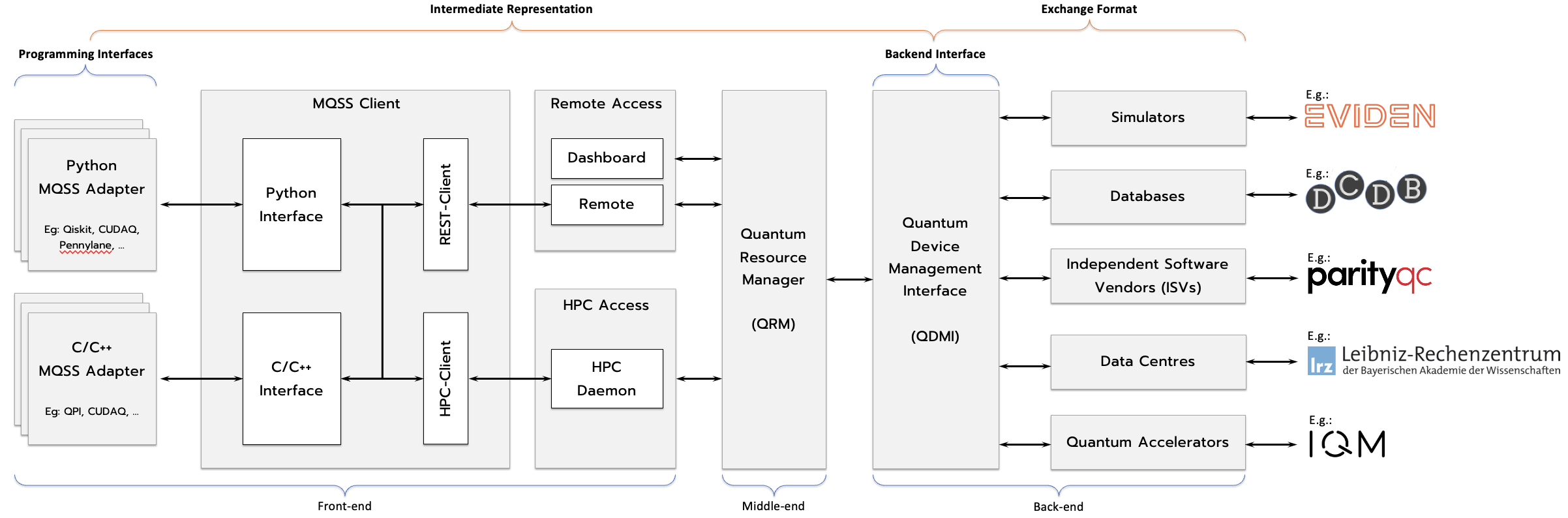}
    \caption{The Munich Quantum Software Stack (MQSS) and its main components: Adapters, such as Qiskit, CUDAQ, Pennylane, and its native C-based QPI, submit a Hamiltonian description as well as both gate- and pulse-level tasks to the client. This client is responsible for automatically routing such requests to Quantum Resource Manager (QRM). Meanwhile, QRM operates as a second-level scheduler, incorporating a Just-In-Time (JIT) LLVM-based compiler and multiple support libraries. Quantum Device Management Interface (QDMI) delivers device specifications to both the QRM during JIT compilation and to the adapters via the client during runtime execution. }
    \Description{
    The The Munich Quantum Software Stack (MQSS) and its main components: Adapters, such as Qiskit, CUDAQ, Pennylane, and its native C-based QPI, are designed to submit a Hamiltonian description as well as both gate- and pulse-level tasks to the client. This client is responsible for automatically routing such requests to QRM. Meanwhile, QRM operates as a second-level scheduler, incorporating a Just-In-Time (JIT) LLVM-based compiler and multiple support libraries. QDMI, on the other hand, delivers each device’s specifications to both the QRM during JIT compilation and to the adapters via the client during runtime execution. 
    }
    \label{fig:mqss}
\end{figure*}

At \lrz\, in collaboration with TU Munich (TUM) and under the Munich Quantum Valley (MQV), these architectural requirements are realized through a custom software stack known as the The Munich Quantum Software Stack (MQSS) \cite{burgholzer2025mqss}.
As shown in Fig.~\ref{fig:mqss}, this software stack 
supports both remote submissions via a REST API and tightly-coupled in-HPC execution, transparently managed by its client. Without requiring any code modifications from the user, the client automatically detects whether a job originates inside or outside an HPC environment and routes it accordingly to the appropriate interface, whether the REST-client for asynchronous access or the HPC-client for local, accelerator-style submission. By design, the software stack facilitates seamless compatibility with various high-level quantum programming frameworks using modular Adapters for frameworks such as CUDAQ, Qiskit, Pennylane, and its own Quantum Programming Interface (QPI) \cite{mete}. By supporting multiple programming interfaces, the software stack allows users to leverage the diverse set of features offered on the client-side by each of these frameworks.

The compilation backend is built on a flexible, Multi-Level Intermediate Representation (MLIR)-based framework capable of supporting multiple dialects, including NVIDIA’s Quake \cite{quake} and Xanadu’s Catalyst \cite{catalyst}. This dialect-agnostic compiler progressively lowers high-level programs into a shared IR, such as the Quantum Intermediate Representation (QIR) \cite{qir}, and finally into hardware-specific instructions. The evolving compiler infrastructure enables integration of additional dialects responding to hardware variability, extended HPC+QC support for hybrid algorithms, and emerging quantum systems.

A QDMI interface \cite{wille2024qdmi}, a lightweight header-only C interface, allows to bridge hardware-specific performance data and the compiler’s optimization flow. Specifically, QDMI enables software tools to query backend-specific metrics, including topology, gate fidelities, noise characteristics, and resource constraints, at runtime, thereby enabling JIT adaptation of compilation and scheduling strategies per platform. It is particularly beneficial to users as just-in-time quantum circuit transpilation can reduce noise \cite{justintime}. 

Practical experience from running hybrid workloads shows that MQSS serves as  a unified platform for integrating quantum accelerators into HPC environments. A key lesson obtained from development is that efficient and successful HPC+QC integration is more about ensuring a seamless interaction between the individual components within the software stack and allowing multiple abstraction layers to target different user bases. In MQSS, this was achieved with the client to allow users to submit jobs seamlessly regardless of the access paths (HPC or remote access), using an MLIR-based multi-dialect compiler to enable efficient circuit compilation techniques, and adaptive backend-awareness via QDMI 
adjusting dynamically to the selected device’s status. 

%% file: src/03_daily_ops.tex
\section{Daily Operations}
 \label{03_daily_ops}

The regular operations of a superconducting quantum computer in an HPC center are different from those of classical HPC nodes. Compared to classical compute hardware which is stable for years, the physical characteristics of superconducting qubits such as their quality metrics (i.e. qubit fidelities) can change over timescales of hours to days. 

For a quantum computer to be a viable component within an HPC environment, its operational characteristics must align with the core tenets of high-performance computing: maximizing uptime, ensuring resource availability and throughput, and enabling efficient, non-blocking workflows. This means, active monitoring, servicing and (re-)calibration is needed to ensure the system delivers reproducible and consistent results for HPC users.

\subsection{Monitoring}
\label{monitoring_chapter}

Due to the dynamic nature of quantum computers an effective integration within HPC environments requires robust monitoring beyond operational uptime. Here, such monitoring is implemented through the Data Center Data Base (DCDB) \cite{netti2020dcdb}, an open-source, plugin-based system designed for continuous and holistic collection of operational and environmental metrics, ranging from sensor data from the cryogenic system and power consumption to performance indicators across compute nodes and other infrastructure. DCDB aggregates this data in a distributed noSQL data store, enabling cross-system correlation and laying the foundation for advanced operational analytics and automated control.

\begin{figure}[ht]
 \centering
 \includegraphics[width=\columnwidth]{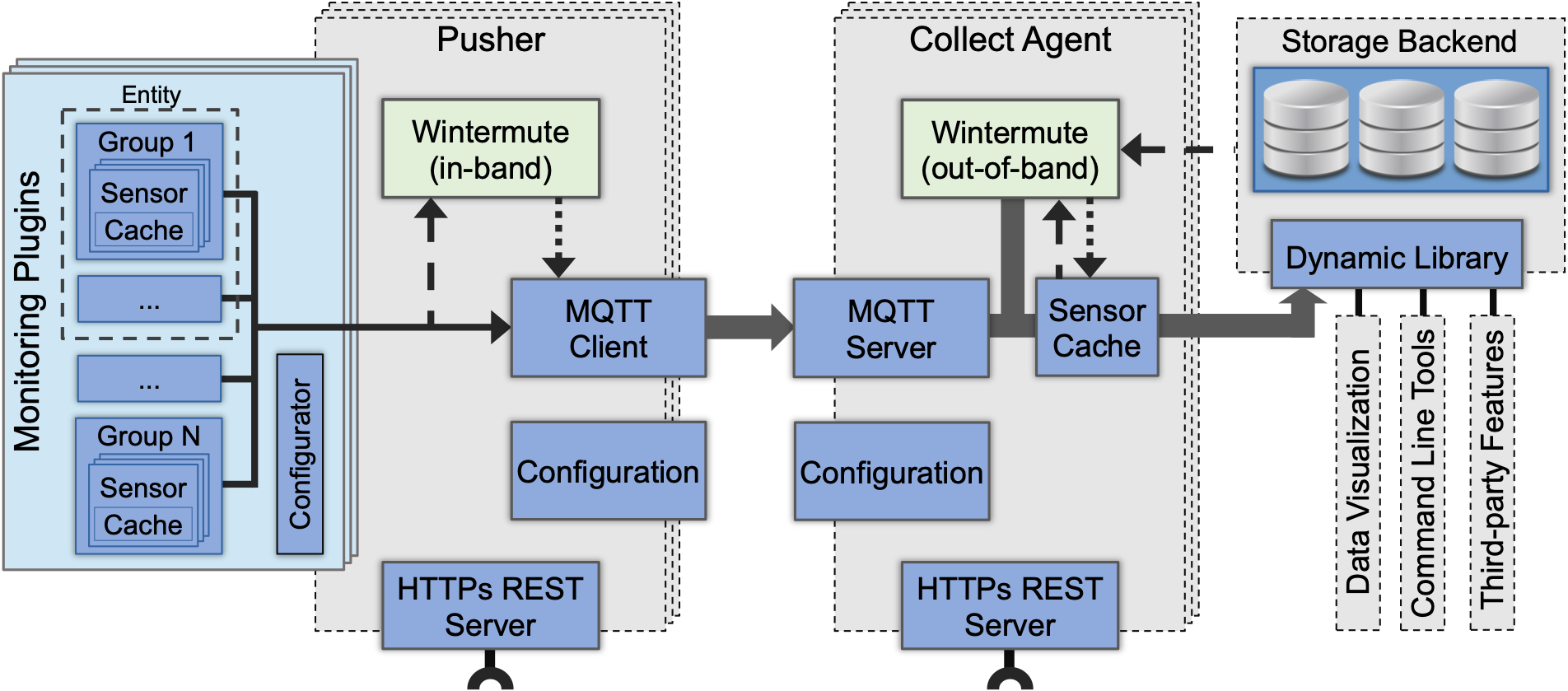}
 \caption{DCDB integration for real-time telemetry-aware quantum execution. It uses the QDMI specification to standardize queries about device properties, constraints, and runtime telemetry data. 
This setup allows to consume these live data during tasks such as JIT compilation and environment-aware optimizations.}
 \label{fig:monitoring}
 \Description{}
\end{figure}

Building upon the infrastructure shown in Fig.~\ref{fig:monitoring}, a QDMI Device has been developed that interfaces with DCDB to acquire telemetry from quantum hardware and its operational environment. This integration uses fine-grained real-time data, for example, qubit fidelities, temperature variations, and noise characteristics. One essential reason is to monitor the computational accuracy of the system, and attempt to identify when a (re-)calibration is required. Another reason is that some error reduction methods rely on the collected information. 

During the project there was increased interest from end users and external software providers to access this telemetry data. Therefore, it was important that the setup enables transparent dissemination of the data so that users and automated tools can access this information without altering workflows.

\subsection{Calibration}
\label{calibration_chapter}

Classical computing processors, Central Processing Units (CPUs) and Graphical Processing Units (GPUs), are fabricated in silicon and their individual properties remain stable following any tuning (i.e. clock adjustment) performed at the factory by the hardware vendor. Qubits in quantum computers -- across all modalities -- are part of dynamic systems that require regular tuning to achieve the best possible computational performance. This process is known as calibration. 

As described in monitoring, the setup regularly runs a suite of algorithmic benchmarks to check the system state. Standardized algorithms such as GHZ state creations are regularly run on all qubits of the QPU or subsets of them. This provides a practical measure of the system's "live" performance and can be used to confirm that the system is ready for user projects. Deviating results can be a sign that a recalibration is needed. 

\begin{figure*}[ht]
    \centering
    \includegraphics[width=\textwidth]{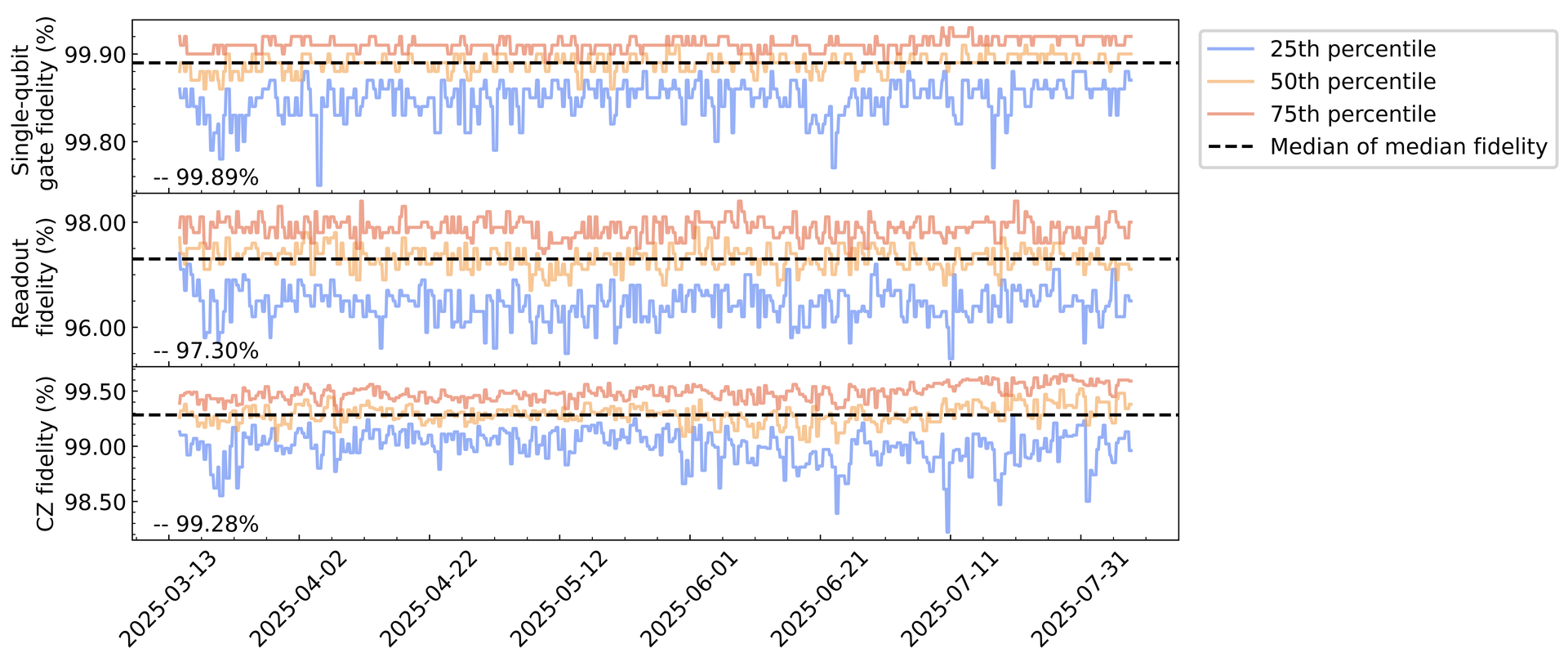}
    \caption{Autonomous calibration performance over 146 days on the quantum computing system, showing consistent single-qubit gate fidelity, readout fidelity and CZ fidelity (two-qubit gate) over time.}
    \Description{Autonomous calibration performance over 146 days on the quantum computing system, showing consistent single-qubit gate fidelity, readout fidelity and CZ fidelity (two-qubit gate) over time.}
    \label{fig:calibration_image}
\end{figure*}

The 20-qubit superconducting quantum computer operates with a fully automated routine recalibration process that requires no human intervention. The system performs a comprehensive full recalibration procedure with the exact timing controlled by the HPC center to optimize operational schedules \cite{calibrating2024, Deng2025}. Operators have the flexibility to choose between quick and full recalibration procedures depending on performance requirements—while quick recalibration offers faster turnaround times (40 minutes), it generally results in lower system performance, whereas the full recalibration procedure (100 minutes), though slower, yields optimal system performance. This automated approach has proven highly reliable with the system operating more than 100 days of continuous operation without human intervention in calibration (see figure \ref{fig:calibration_image}).

With the workloads that have been seen since the inauguration of the system, it showed it is critical that the center retains full control over scheduling these maintenance and calibration slots to align with current and upcoming user workloads. 

\subsection{On-Site Support}
As a solid-state platform, the on-site support requirements of superconducting quantum computers differ from those of other modalities. The chip-based architecture eliminates the need for certain types of manual intervention, such as the physical cleaning or realignment of optical components that can be necessary for trapped-ion systems or neutral atoms. 

Consequently, during normal operations, the need to physically touch the system is limited to adding approximately ten liters of liquid nitrogen every week to the cryogenic cooling system. Most other parts of the system can be controlled and debugged remotely. For the communication between the vendor and the HPC center, the use of a ticketing system has been deemed helpful. A knowledge base showing videos of regular tasks was also provided and deemed helpful. In rare cases, on-site support is needed if the control electronics or the cryostat system has a hardware issue leading to part replacement or in-depth debugging. 

\subsection{Periodic Maintenance}
To ensure long-term reliability and optimal performance, the operational model includes periodic scheduled maintenance windows. These planned activities are coordinated with the HPC center's schedule to minimize disruption to user access.

A one-day preventive maintenance procedure is performed approximately every six months. This activity involves flushing the liquid nitrogen system. The purpose of this flush is to remove any ice or other debris that might have accumulated from ambient air during previous procedures. Other activities that are performed in some of these preventive maintenance windows based on the age of the systems include checking or replacing the batteries for the Uninterruptible Power Suppy (UPS), or replacing tip seals of the cryogenic system's pumps. The preventive maintenance procedure can be extended to a longer maintenance window that is used to perform more significant system upgrades, such as deploying major updates to the control software and firmware.

\subsection{Recovering from outages}

HPC nodes can typically be restarted with relative ease following a power or cooling failure. Quantum computers, on the other hand, require a more involved recovery process, since loss of power or cooling can cause the QPU to warm up above its normal 10 mK operating temperature. Unplanned warm up events can also be caused by the cooling water temperature exceeding the upper temperature limit, triggering a shutdown of the cryogenic pumps.

Once active cooling ceases, the calibration state of the QPU can be largely maintained if the QPU temperature remains below 1K. In practice, it takes two minutes to exceed this temperature after a fault in the cooling system. For small temperature excursions that stay below 1K, calibration can often be restored by the automated calibration system without requiring manual intervention. However, for higher temperatures, a full calibration is necessary.

Another rare, potential issue is the loss of vacuum inside the cryostat, which can expose the QPU to ambient air resulting in oxidation. Fortunately, the vacuum integrity of the system is typically maintained during outages for several weeks, unless the system is deliberately opened or physically moved.

Restarting a quantum computer after a major outage involves several sequential steps. First, the underlying issue causing the outage must be identified and resolved. Once the issue is addressed, the cryostat must be cooled down to its operating temperature, a process that can take from two to five days depending on the thermal mass of the cryostat and the temperature reached during the outage. Once the system is below 100 mK, and ideally back to 10 mK, recalibration and benchmark verification of the system can occur. This recovery period of two to five days caused by the cryostat cooldown highlights the benefits of integrating redundant power and cooling water supplies.

%% file: src/04_onboarding.tex
\section{Experience in onboarding the first users}
\label{04_onboarding}

Providing hardware access is only the first step; the value of a quantum computer within an HPC center is ultimately realized through the scientific output of its users. Our experience shows that a structured onboarding and training program is essential to bridge the significant conceptual and practical gaps between classical HPC and quantum computing. We identified two distinct user groups with different training needs: expert quantum users requiring device-specific knowledge as well as HPC knowledge and traditional HPC users needing a fundamental introduction to quantum computing principles.

First, a list of candidate users for the early user phase of the quantum computer was compiled. These candidates were selected through a review process based on the relevance of their research topics, the clearly articulated scientific plan to build a workflow combining  quantum and HPC resources, and the likelihood of delivering meaningful results within the given timeline (early users were required to submit a final report). Candidates were also considered based on their prior collaborations with \lrz, which helped streamline communication through existing channels, and institutional affiliation to the MQV.

The quantum computing vendor provided training to a cohort of internal users at \lrz. The primary challenge for the cohort was not understanding quantum algorithms, but learning how to adapt their workflows to get the most out of the specific superconducting hardware. Therefore, the training focused on device-specific characteristics, such as optimizing for the QPU topology, the native gate set, and noise profiles. Through an interactive training session, users were taught "tips and tricks" for circuit compilation and how to implement error mitigation methods tailored to the machine, enabling them to optimize their research projects for the available hardware. Hands-on exercises were provided with Jupyter notebooks. This interactive format allowed participants to gain practical experience writing and executing quantum circuits on the system while providing scaffolding through the Use-Modify-Create model \cite{lee2011computational} – a concept that promotes a structured path from guided use, through experimental modification, to independent creation that has also been used in adult education \cite{lao2019deep}.


To support the implementation of the planned research projects, we introduced a structured mentorship model. The mentorship model was led by LRZ solution architects, chosen based on their domain expertise and existing relationships with early users. Early user training began with quantum circuit submissions to a digital twin of the quantum computer (an emulator). A second crucial part of the training was learning how to submit jobs to HPC resources within the existing workflow framework. These hands-on sessions, centered on Jupyter notebook exercises, enabled users to test algorithms in advance and become familiar with the system. Finally, “open-mic” sessions were hosted where early users and solution architects could exchange real-time feedback, report issues, and work together on implementing immediate solutions. This close partnership accelerated the transition from concept to execution and provided the operations team with direct insights into user needs and common obstacles.

Based on the feedback collected, we compiled a set of Frequently Asked Questions (FAQ). The questions were organized into categories: Getting Started; Job Submission \& Execution; Job Tracking \& Results; System \& Hardware Information; Resource Usage; and Budgeting. This organization not only made information easier to find, but also helped us identify and prioritize user issues. For example, many users found it difficult to navigate large job histories on the dashboard, which led us to implement more efficient pagination in the results section. The FAQ process also revealed previously unanticipated needs. Users requested features such as batch-job support, access to qubit coupling maps, more robust job restart tools after system outages, and greater transparency in the qunatum circuit compilation process. Additionally, some users needed pulse-level access, enabling them to move beyond circuit-based programming and design hardware-specific control sequences.


The initial user projects have already led to first publications and preprints (e.g. \cite{bentellis,bickley}), confirming that a dedicated onboarding program is effective in converting hardware access into tangible scientific output.

%% file: src/05_conclusion.tex
\section{Conclusion}
\label{05_conclusion}

The practical integration of the 20-qubit superconducting quantum computer into \lrz's HPC infrastructure demonstrates that superconducting on-premise quantum systems can be effectively co-located with classical HPC resources when their unique operational requirements are systematically addressed. From site selection through sustained operations, our experience validates and contextualizes four key lessons:
\begin{enumerate}
    \item \textbf{Stricter but manageable facility requirements} – As discussed in Section \ref{02_space}, superconducting quantum computers impose more stringent environmental constraints than classical HPC nodes, including stricter requirements on vibrations, background electromagnetic noise, and room temperature stability ($\Delta T < 1$ °C over 24 h). The site survey methodology proved essential in quantifying and mitigating environmental noise, ensuring stable system performance. These findings confirm that, with a rigorous pre-installation assessment, HPC facilities can meet quantum hardware requirements without prohibitive modifications.
    
    \item \textbf{Quantum Computers are dynamic systems requiring regular recalibration} – Sections \ref{monitoring_chapter} and \ref{calibration_chapter} detail how automated, scheduler-controlled calibration routines allow the quantum computer to maintain consistent gate and readout fidelities over more than 100 days of unattended operation. Controlling calibration scheduling through the HPC resource management framework offers flexibility to users with large computational workloads.

    \item \textbf{Necessity of redundant infrastructure} – Recovery procedures described in Section 3.6 highlight the disproportionate downtime caused by cooling or power interruptions, with worst case scenarios resulting in cryostat cooldown periods of up to ten days and the need for a full calibration after cooldown. The presence of redundant cooling water and uninterruptible power supplies mitigates these risks.

    \item \textbf{Importance of structured onboarding and training} – Section \ref{04_onboarding} demonstrates that a targeted onboarding program, combining hardware-specific training, mentorship, and iterative feedback mechanisms such as open-mic sessions, accelerated user readiness and yielded early scientific results. The segmentation of user groups and adoption of the Use–Modify–Create model ensured that both quantum specialists and HPC practitioners could productively access the system.

\end{enumerate}

The practical experiences gained in this case study demonstrate that HPC+QC integration succeeds when quantum systems are treated as first-class citizens within the HPC ecosystem. By combining rigorous facility preparation, automated calibration, redundant infrastructure for power and cooling, adaptive software layers, and structured user onboarding, quantum resources can be handled like any other HPC resource following established data center processes.

%% file: src/06_acks.tex
\begin{acks}
This work is supported by the German Federal Ministry of Research, Technology and Space (BMFTR) (grant 13N16063, Q-Exa and  grant 13N16690, Euro-Q-Exa) and the Bavarian State Ministry of Science and the Arts through funding, as part of MQV, Q-DESSI. 
The authors would also like to thank the Q-Exa project team, including  LRZ’s  Systems and Labs (SL) group, User Enablement and Applications (UEA) group, and the Quantum Integration Software (QIS) group, as well as IQM Quantum Computers for their support and collaboration throughout the integration process.
\end{acks}
